\documentclass{revtex4}
\usepackage{amsthm,amssymb,amsmath}				
\usepackage{graphicx,comment}
\usepackage{float}   
 \usepackage{xcolor}
 \usepackage{color}
 \definecolor{mycolor}{rgb}{0.8, 0.2, 0.2} 
 \usepackage{hyperref}

\begin{document}
%\fontsize{8}{9}\selectfont
%\preprint{APS/123-QED}

\title{Klein-Gordon particles in a nonuniform external magnetic field in Bonnor-Melvin rainbow gravity background}

\author{Omar Mustafa}
\email{omar.mustafa@emu.edu.tr (Corr. Author)}
\affiliation{Department of Physics, Eastern Mediterranean University, 99628, G. Magusa, north Cyprus, Mersin 10 - Türkiye}

\author{Abdullah Guvendi}
\email{abdullah.guvendi@erzurum.edu.tr}
\affiliation{Department of Basic Sciences, Erzurum Technical University, 25050, Erzurum, Türkiye}

\date{\today}

\begin{abstract}
\vspace{0.15cm}
\setlength{\parindent}{0pt}

{\footnotesize We investigate the effect of rainbow gravity on Klein-Gordon (KG) bosons in a quantized nonuniform magnetic field in the background of Bonnor-Melvin (BM) spacetime with a cosmological constant. In the process, we show that the BM spacetime introduces domain walls (i.e., infinitely impenetrable hard walls) at \(r = 0\) and \(r = \pi/\sqrt{2\Lambda}\), as a consequence of the effective gravitational potential field generated by such a magnetized BM spacetime. As a result, the motion of KG particles/antiparticles is restricted indefinitely within the range \(r \in [0, \pi/\sqrt{2\Lambda}]\), and the particles and antiparticles cannot be found elsewhere. Next, we provide a conditionally exact solution in the form of the general Heun function \(H_G(a, q, \alpha, \beta, \gamma, \delta, z)\). Within the BM domain walls and under the condition of exact solvability, we study the effects of rainbow gravity on KG bosonic fields in a quantized nonuniform external magnetic field in the BM spacetime background. We use three pairs of rainbow functions: \( f(u) = (1 - \tilde{\beta} |E|)^{-1}, \, h(u) = 1 \); and \( f(u) = 1, \, h(u) = \sqrt{1 - \tilde{\beta} |E|^\upsilon} \), with \(\upsilon = 1,2\), where \(u = |E| / E_p\), \(\tilde{\beta} = \beta / E_p\), and \(\beta\) is the rainbow parameter. We find that such pairs of rainbow functions, \((f(u), h(u))\), fully comply with the theory of rainbow gravity, ensuring that \(E_p\) is the maximum possible energy for particles and antiparticles alike. Moreover, we show that the corresponding bosonic states form magnetized, rotating vortices, as  intriguing consequences of such a magnetized BM spacetime background.} 

\end{abstract}


\maketitle

%\tableofcontents

\section{Introduction}

%\begin{figure*}[ht!]
%   \centering
%   \includegraphics[scale=0.60]{.....   .eps}\\
%   \caption{ \footnotesize }\label{fig:1}
%\end{figure*}
\vspace{0.15cm}
\setlength{\parindent}{0pt}

Magnetic fields play a fundamental role in a vast array of astrophysical phenomena, ranging from stellar structures and accretion disks to galactic cores and the intergalactic medium. The presence of these fields in regions dominated by intense gravitational forces, particularly near compact massive objects, has driven extensive research within the framework of general relativity. Among the exact solutions to the Einstein-Maxwell equations, the BM universe offers a compelling scenario, describing a static, cylindrically symmetric magnetic field aligned with its symmetry axis and immersed in its own gravitational field \cite{R1.1,R1.2,R1.3}. In this model, the magnetic field contributes to the energy-momentum tensor, thereby curving the surrounding spacetime. Its intensity diminishes with radial distance to prevent gravitational collapse and maintain stability. To sustain equilibrium and ensure a homogeneous field, a nonzero positive cosmological constant is incorporated into the metric. Consequently, the BM spacetime, incorporating a positive cosmological constant \(\Lambda > 0\), is described by \cite{R1.1}:  
\begin{equation}
ds^{2} = -dt^{2} + dr^{2} + \alpha^{2} \sin^{2} \left( \sqrt{2\Lambda} r \right) d\varphi^{2} + dz^{2},  
\label{I.1}
\end{equation}  
where \(z\in[-\infty,\infty],\,\varphi\in[0,\pi],\,r\in[0,\infty]\), and the cosmological constant \(\Lambda\) has the units of inverse length squared. Moreover, the corresponding magnetic field is given by \(H = \sqrt{\Lambda} \, \alpha \, \sin \left( \sqrt{2\Lambda} r \right)\). Where, \(\alpha\) is an integration constant \cite{R1.1}, linked to the cosmic string parameter, which dictates the deficit angle of the conical spacetime, constrained by \(0<\alpha^2=1-\eta/2\pi<1\), with \(\eta\) representing the linear mass density of the cosmic string \cite{R1.4,R1.4.1}. Notably, Zofka \cite{R1.1} demonstrated that at \(\sqrt{2\Lambda }r=\pi\), the circumferences of constant-\(r\) rings vanish, implying the presence of an axial characteristic. In the present study, we rigorously establish that this location coincides precisely with an infinite hard wall (domain wall), which is recognized as one of the fundamental topological defects-while examining the behavior of KG test particles and antiparticles within the BM spacetime that incorporates a cosmological constant. 

\vspace{0.15cm}
\setlength{\parindent}{0pt}

The grand unified theory predicts the emergence of topological defects in spacetime \cite{R1.5,R1.6,R1.7,R1.8}, including domain walls \cite{R1.6,R1.7}. These defects have profound implications for quantum mechanical particles, influencing their spectral characteristics and dynamical behavior \cite{R1.9,R1.10,R1.11,R1.12,R1.13,R1.14,R1.15,R1.16,R1.17,R1.18,AH,AY,AO,R1.19,R1.20,R1.21}. Investigating the relationship between gravitational fields and quantum systems remains a key motivation for exploring various quantum-mechanical scenarios in the context of different spacetime structures. The BM spacetime \cite{R1.1,R1.2,R1.3,R1.22,R1.23,AO-2,AAS} is one of the intriguing frameworks for this pursuit.

\vspace{0.15cm}
\setlength{\parindent}{0pt}

In contrast, within the framework of rainbow gravity (RG), the energy of a test particle is known to influence the underlying spacetime at extremely high energies (i.e., the ultraviolet regime), leading to modifications in the standard relativistic energy-momentum dispersion relation (MDR) \cite{R1.24,R1.25,R1.26,R1.27,R1.28,R1.29,R1.30}. This modification is captured as follows:
\begin{equation}
E^{2}f\left(u\right) ^{2}-p^{2}h(u)
^{2}=m_\circ^{2}, \label{I.2}
\end{equation}
which emerges as a fundamental prediction of several quantum gravity paradigms, including string field theory, loop quantum gravity, and non-commutative geometry \cite{R1.31,R1.32,R1.33}. Accordingly, the BM spacetime undergoes a transformation under the influence of RG, yielding the modified metric:
\begin{equation}
ds^{2}=-\frac{dt^{2}}{f\left( u\right) ^{2}}+\frac{1}{h\left( u\right) ^{2}}\left(dr^{2}+\alpha ^{2}\sin ^{2}\left( \sqrt{2\Lambda }r\right)
\,d\varphi ^{2}+dz^{2}\right),  \label{I.3}
\end{equation}
where \(f(u)\) and \(h(u)\) are the so-called rainbow functions, which satisfy the conditions \(\lim\limits_{u\rightarrow 0}f\left( u\right) =1=\lim\limits_{u\rightarrow 0}h\left( u\right)\). The parameter \(u=|E|/E_{p}\) ranges from 0 to 1, ensuring that the standard energy-momentum dispersion relation is retrieved in the infrared regime. It should be noted that \(u\) is a finely tuned parameter that enables RG to influence both relativistic particles and antiparticles alike \cite{R1.30,R1.34,R1.35}. Moreover, the choice of rainbow functions ensures that the Planck energy \(E_p\) serves as a natural boundary distinguishing quantum and classical physics while introducing an additional invariant alongside the speed of light.   

\vspace{0.15cm}
\setlength{\parindent}{0pt}

Our primary objective, in the current study, is to derive a conditionally/quasi-exact solution with rigorously imposed boundary conditions for KG quantum particles and antiparticles in an external nonuniform magnetic field in BM spacetime background, and within the framework of RG  (\ref{I.3}). The presence of the sinusoidal term \(\sin(\sqrt{2\Lambda}r)^2\) in the BM-spacetime metric (\ref{I.3}) inherently suggests that \(\sin(\sqrt{2\Lambda}r)^2\in[0,1]\). This would in turn impose some constraints on the allowed range of the radial coordinate \(r\), which requires that \(\sqrt{2\Lambda}r=\tau\pi\) for \(\tau\in\mathbb{Z}\), with \(\tau=0,1,2,\dots\). Setting the radial coordinate origin at \(r=0\) (i.e. \(\tau=0\)) naturally implies that the upper bound of \(r\) occurs at \(r=\pi/\sqrt{2\Lambda}\) when \(\tau=1\), which means that the allowed domain for \(r\) is \(r\in[0,\pi/\sqrt{2\Lambda}]\). The upper bound indicates that the system's dynamics is profoundly determined by the cosmological constant. This interpretation aligns with \v{Z}ofka’s \cite{R1.1} proposition that condition \(\sqrt{2\Lambda}r=\pi\) signifies the presence of a fundamental boundary or axis. To the best of our knowledge, the model problem under consideration in the current study has never been attempted before and/or reported elsewhere.

\vspace{0.15cm}
\setlength{\parindent}{0pt}

In this study, we conduct a comprehensive investigation into the impact of RG on the dynamics of KG particles interacting with Bonnor-Melvin (BM) domain walls in the presence of an external magnetic field. Our analysis explores how modified energy-momentum dispersion relations influence the behavior of these particles/antiparticles within the curved spacetime framework, shedding light on potential deviations from conventional relativistic predictions. In Section \ref{sec:2}, we demonstrate that the boundaries at \(\sqrt{2\Lambda}r=0,\text{and }\sqrt{2\Lambda}r=\pi\) correspond to two domain walls, which act as infinite impenetrable boundaries, confining the motion of particles within the interval \(r \in [0, \pi/\sqrt{2\Lambda}]\). This conclusion arises from an analysis of the effective potential induced by the gravitational field of the BM spacetime. Notably, we avoid the widely used approximation \(r \ll 1 \Rightarrow \sin(r) \approx r \Rightarrow \tan(r) \approx r\) (e.g. \cite{R1.4,R1.4.1}), as we believe that this simplification distorts the quantum mechanical dynamics of the system and overlooks the characteristic domain walls inherent to BM spacetime. Consequently, we derive a non-perturbative wave equation in this section. Section \ref{sec:2A}, investigates the impact of RG by incorporating two distinct pairs of rainbow functions. In Section \ref{sec:3}, we use two sets of rainbow functions: (i) the Magueijo-Smolin pair, \( f\left(u \right) =1/\left( 1-\tilde{\beta}\left\vert E\right\vert \right) \), \( h\left( u \right) =1 \), derived from the varying speed of light hypothesis \cite{R1.36}, and (ii) the set inspired by loop quantum gravity \cite{R1.38,R1.39}, namely \(f(u) = 1\) and \(h(u) = \sqrt{1-\tilde{\beta}|E|^\upsilon}\), with \(\upsilon = 1, 2\). These two sets are known to align fully with the RG principles, enforcing an upper energy limit at the Planck scale \(E_p\), which ensures that neither particles nor antiparticles can exceed this threshold (e.g. \cite{R1.34,R1.35,R1.40}). Finally, in Section \ref{sec:4}, we present our concluding remarks.

\section{The wave equation and the effective gravitational potential field}\label{sec:2}

In this section, we derive a non-perturbative wave equation by incorporating RG effects. To begin, we rescale the coordinates in equation (\ref{I.3}), following the approach in \cite{R1.4}, so that \(\sqrt{2\Lambda}r \rightarrow \rho\) and \(\sqrt{2\Lambda}\varphi \rightarrow \varphi\). This transformation allows us to express the BM-spacetime in the RG metric form as
\begin{equation}
ds^{2}=-\frac{dt^{2}}{f\left( u\right) ^{2}}+\frac{1}{h\left( u\right) ^{2}}\left(dz^{2}+\frac{1}{2\Lambda}\left[d\rho^{2}+\alpha ^{2}\sin ^{2}\left( \rho\right)
\,d\varphi ^{2}\right]\right).  \label{II.1}
\end{equation}
\begin{figure*}[ht!]
\centering
\includegraphics[width=0.25\textwidth]{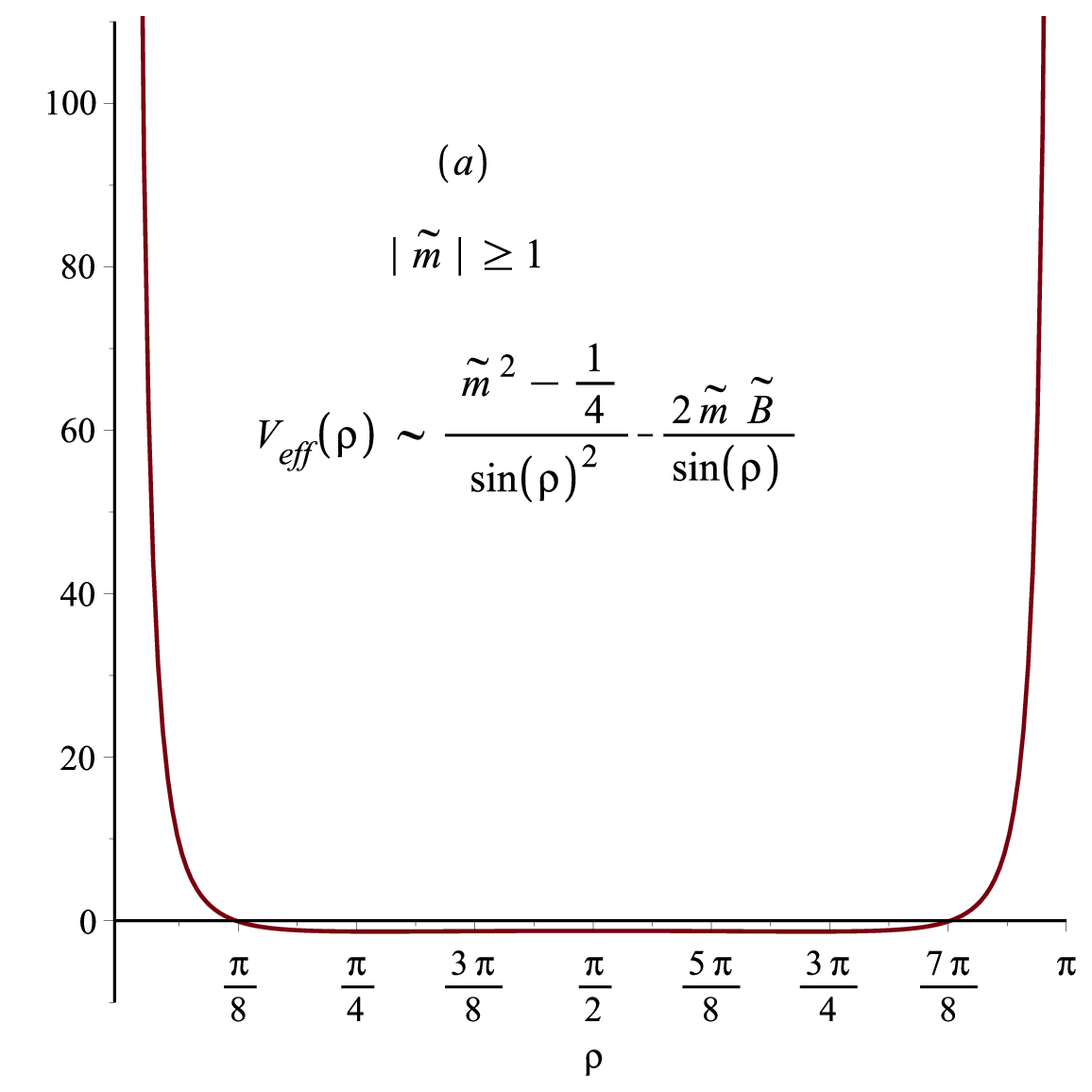}
\includegraphics[width=0.25\textwidth]{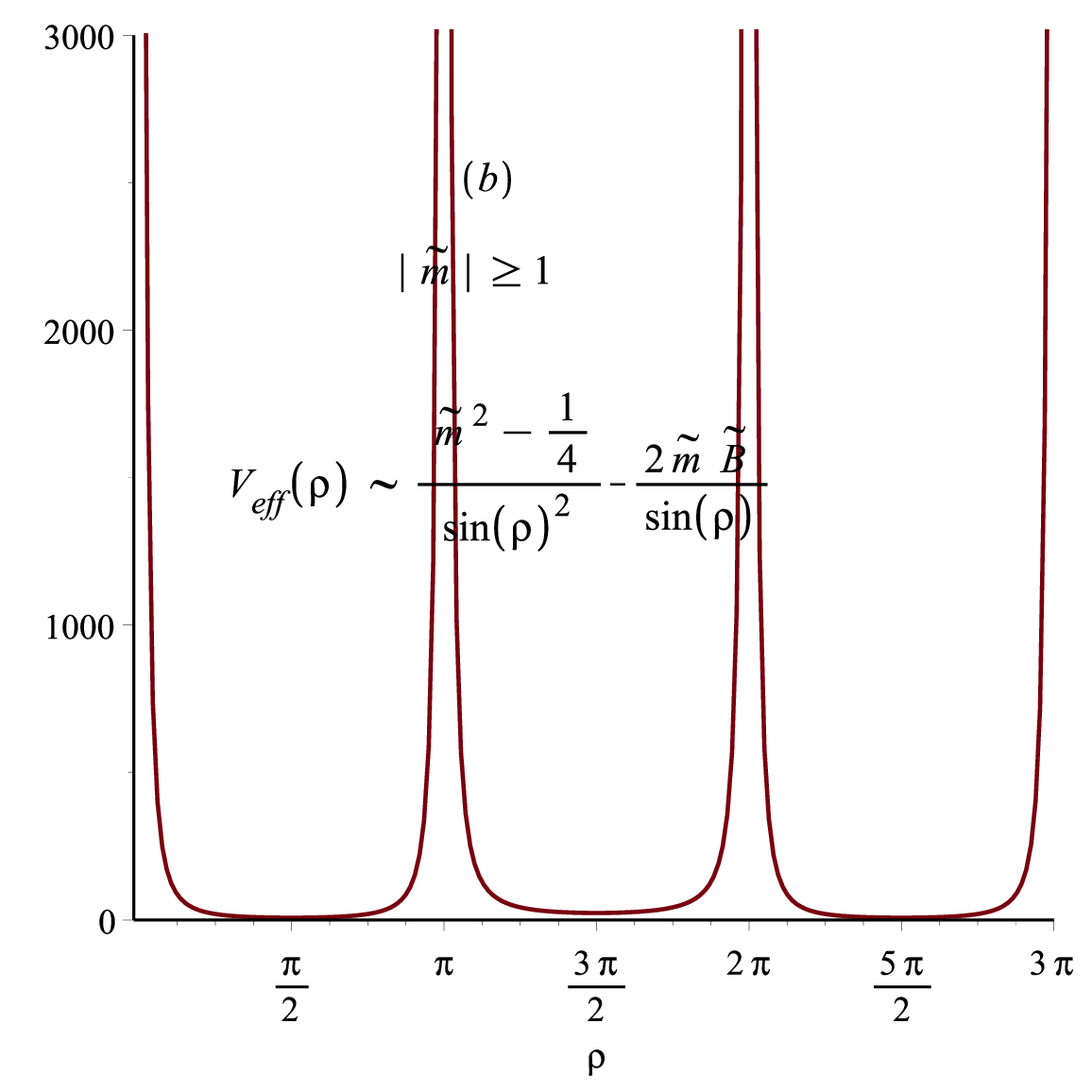}
\includegraphics[width=0.25\textwidth]{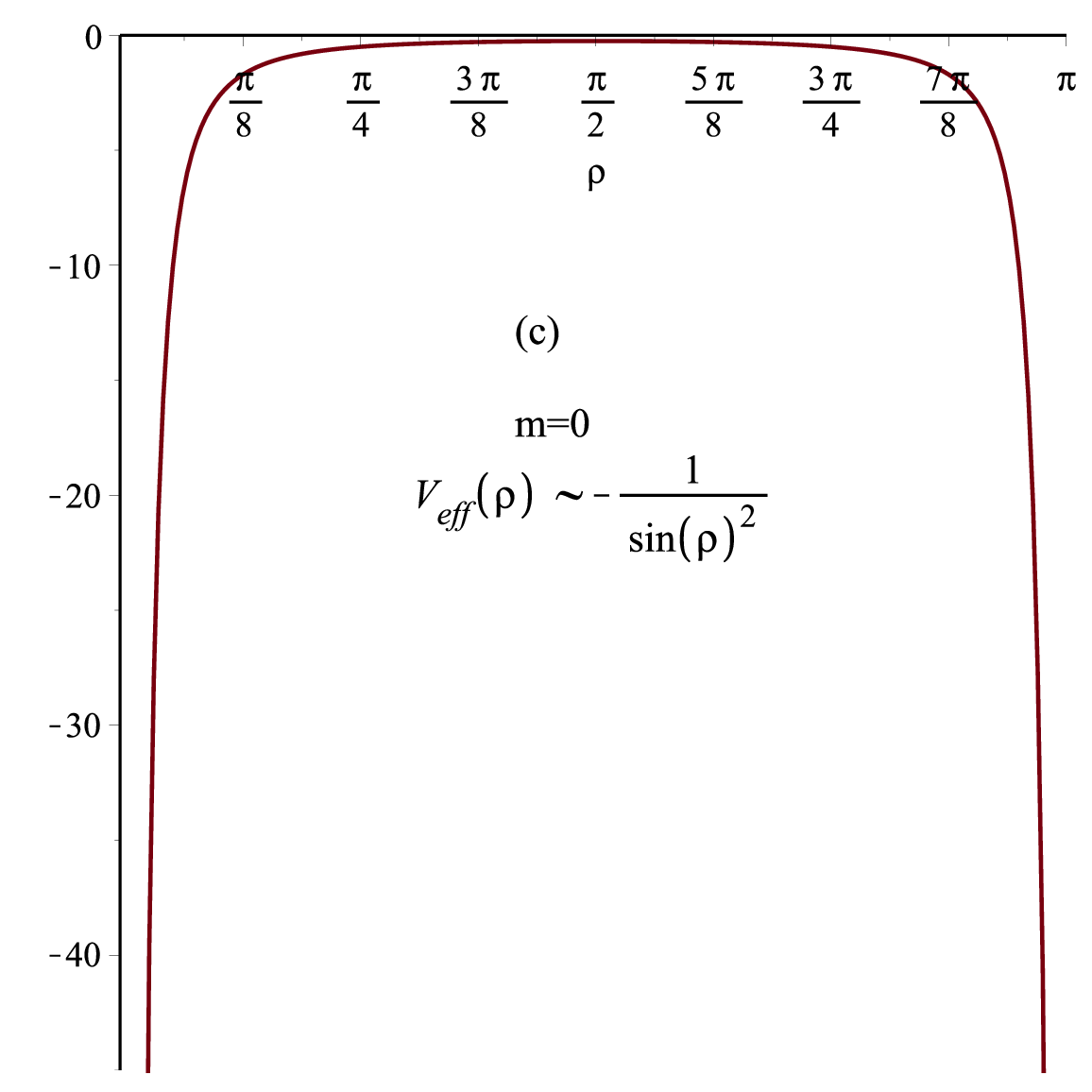}
\caption{\footnotesize  The effective potential (\ref{II.7.1}) for the magnetic quantum number \(|m| \geq 1\): (a) for \(r \in [0, \pi]\), (b) for \(r \in [0, 6\pi]\), and (c) for \(m = 0\).}
\label{fig1}
\end{figure*}
For this spacetime, the contravariant metric tensor \(g^{\mu\nu}\) has the non-vanishing elements
\begin{equation}
    g^{00}=-f(u)^2,\,\,g^{11}=2\Lambda h(u)^2,\,\,g^{22}=\frac{2\Lambda h(u)^2}{\alpha^2\sin(\rho)^2},\,\,g^{33}=h(u)^2, \label{II.2}
\end{equation}
and the determinant of the covariant metric tensor is given by \(\text{det}(g_{\mu\nu})=g=-\frac{\alpha^2 \sin(\rho)^2}{ 4\Lambda^2 h(u)^6 f(u)^2}\). Next, we describe a KG particle/antiparticle in an external magnetic field in the BM spacetime using the equation
\begin{equation}
    \frac{1}{\sqrt{-g}}\left(\partial_\mu -ieA_\mu\right)\sqrt{-g}\,g^{\mu\nu} \,\left(\partial_\nu-ieA_\nu \right)\, \psi(t,\rho,\varphi,z)=m_\circ ^2\, \psi(t,\rho,\varphi,z),\label{II.3}
\end{equation}
where \(m_\circ\) represents the rest mass energy of the particle. We now introduce the substitutions
\begin{equation*}
A_{\varphi}=\frac{\alpha B_{\circ}}{\sqrt{2\Lambda}} \sin(\rho),\quad \psi(t,\rho,\varphi,z)=e^{i(m\varphi+k\,z-Et)} R(\rho),
\end{equation*}
which lead to the following differential equation for \(R(\rho)\):
\begin{equation}
R ^{\prime \prime }\left( \rho\right) +\frac{1}{\tan \left( \rho\right) }R
^{\prime }\left( \rho\right) +\left( \mathcal{E} -\frac{\left[\tilde{m}-\tilde{B}\,\sin(\rho)\right]^2}{\sin \left(
\rho\right) ^{2}}\right) R\left( \rho\right) =0,  \label{II.4}
\end{equation}
where
\begin{equation}
\tilde{m}=\frac{m}{\alpha },\quad \tilde{B}=\frac{eB_{\circ}}{\sqrt{2\Lambda}},\quad\mathcal{E} =\frac{f\left( u\right) ^{2}E^{2}-m_{\circ}^{2}}{ 2\Lambda h\left( u\right) ^{2}}-\frac{k^{2}}{2\Lambda}. \label{II.5}
\end{equation}
At this point, one should notice that our \(A_{\varphi}\) satisfies the electromagnetic field fundamental invariant \(\frac{F_{\mu\nu}F^{\mu\nu}}{2}=g^{11}g^{22}(\partial_\rho A_{\varphi}-\partial_{\varphi}A_{\rho})^2=||\textbf{B}||^2-||\textbf{E}||^2\) and yields a nonuniform magnetic field \(B_z=\sqrt{2\Lambda}h(u)^2\,B_{\circ}\sqrt{\frac{1}{sin(\rho)^2}-1}\). Hereby, one should observe that RG does not only affect the energies of the particle and antiparticle but also does have its say on the external magnetic field \(B_z\). In \eqref{II.5}, \(m=0,\pm1,\pm2,\cdots\) denotes the magnetic quantum number. Ultimately, we obtain the following equation:
\begin{equation}
    R ^{\prime \prime }\left( \rho\right) +\frac{1}{\tan \left( \rho\right) }R
^{\prime }\left( \rho\right) +\left( \tilde{\mathcal{E}} -\frac{\mathcal{M}^2}{\sin \left(
\rho\right) ^{2}}+\frac{2\tilde{m}\tilde{B}}{\sin \left(
\rho\right)}\right) R\left( \rho\right) =0, \label{II.6}
\end{equation}
where \(\tilde{\mathcal{E}}=\mathcal{E}-\tilde{B}^2\) and \(\mathcal{M}^2=\tilde{m}^2 \Rightarrow \mathcal{M}=|\tilde{m}|\), which distinguishes the effects of \(\tilde{m}=|\tilde{m}|\) (in $\mathcal{M}^2$) and \(\tilde{m}=\pm|\tilde{m}|\) (in $2\tilde{m}\tilde{B}$) on the spectroscopic structure of the problem at hand. At this point, in order to cast this equation in a one-dimensional Schrödinger-like form and find the effective gravitational potential, we make the substitution \(R(\rho)=U(\rho)/\sqrt{\sin(\rho)}\), yielding
\begin{equation}
    U''(\rho)+\left[\tilde{\lambda}-\frac{\left(\mathcal{M}^2-1/4\right)}{\sin(\rho)^2}+\frac{2\tilde{m}\tilde{B}}{\sin \left(\rho\right)}\right]U(\rho)=0, \label{II.7}
\end{equation}
where \(\tilde{\lambda}=\tilde{\mathcal{E}}+1/4\). It is evident that the corresponding effective gravitational potential is
\begin{equation}
V_{eff}(\rho)=\frac{\left(\mathcal{M}^2-1/4\right)}{\sin(\rho)^2}-\frac{2\tilde{m}\tilde{B}}{\sin \left(\rho\right)}. \label{II.7.1}
\end{equation}
This potential exhibits singularities at \(\rho=\tau \pi\), where \(\mathbb{Z}\ni\tau=0,1,2,\cdots\). These singularities are a direct result of the magnetized BM-spacetime background, which manifests as an infinite sequence of standard textbook infinite wells (each of width \(\pi\)), as depicted in Figures \ref{fig1}(a) and \ref{fig1}(b). Each well is represented by two domain walls (domain walls are well-known topological defect in the early universe). As such, the radial coordinate is constrained to the range \(\rho \in [0,\pi]\). Consequently, a quantum particle remains confined between the first two domain walls at \(\rho=0\) and \(\rho=\pi\), unable to penetrate or propagate beyond these walls. Moreover, the quantum mechanical structure of the problem generates a repulsive core for \(\left( \tilde{m}^{2}-1/4\right) >0\), i.e., for \(\left\vert m\right\vert \geq 1\), with a repulsive gravitational force field that intensifies as \(|m|\) increases. For \(m=0\), however, the potential becomes attractive (with an extremely strong attraction for \(r << 1\), as shown in Fig. \ref{fig1}(c)). To account for bound states of KG particles (\(E=E_+=+|E|\)) and antiparticles (\(E=E_-=-|E|\)), one must impose the condition \(\tilde{\lambda}\equiv \tilde{\lambda}(E^2) > 0\). For \(m=0\), however, \(\tilde{\lambda}\equiv \tilde{\lambda}(E^2) < 0\), and as clearly illustrated in Fig. \ref{fig1}(c), the corresponding states are not steady but instead exhibit time-dependent growth or decay, with their modes following the behavior \(\psi \propto e^{-iE t}\), where the decay time is given by \(\tau=\frac{1}{|\Im E|}\). In the current methodological framework, we focus solely on KG particles and antiparticles confined to move within the impenetrable infinite domain walls (\(r \in [0, \pi]\)), generated by the BM-spacetime under an external magnetic field (with \(A_{\varphi}=\frac{B_{\circ}}{2\Lambda} \sin(\rho)\)) and possessing magnetic quantum numbers \(m=\pm 1, \pm 2, \dots\).

\subsection{Conditional exact solvability for the wave equation (\ref{II.6})} \label{sec:2A}

To solve (\ref{II.6}), we use the substitution
\begin{equation}
    R(\rho)=\sin(\rho)^{|\tilde{m}|}\sqrt{1-\sin(\rho)}\,H(\rho) \label{II.7.2}
\end{equation}
followed by the change of variables  \(\rho = \arcsin(x)\) to obtain
\begin{equation}
(x^2-1)\,H''(x) + \left[M_0 \,x+1-\frac{M_1}{x}\right]\,H'(x) + \left[q_{_{1}}+\frac{q_{_{2}}}{x}\right]\,H(x) = 0, \label{II.8}
\end{equation}
where,
\begin{gather}
M_0=2|\tilde{m}|+3,\quad M_1=2|\tilde{m}|+1, \notag \\
q_{_{1}}=\tilde{m}^2-\tilde{\mathcal{E}}+2|\tilde{m}|+\frac{3}{4},\quad 
q_{_{2}}=|\tilde{m}|+\frac{1}{2}-2\,\tilde{m}\tilde{B}.   \label{II.9}
\end{gather}
This equation is then solved using a power series solution of the form
\begin{equation}
 H(x) = \sum_{j=0}^{\infty} A_j (-x)^{j + \sigma}, \label{II.9.1}
\end{equation}
to imply  \(\sigma = 0\) or \(\sigma=-2|\tilde{m}|\). The latter should be discarded as it yields \(H(0)\rightarrow\infty\) that violates the finiteness of the radial wave function at the domain walls. Consequently, one obtains \(A_0 = 1\), \(A_1 = - q_{_{2}}/M_1,\) and the three terms recurrence relation 
\begin{equation}
A_{j+2}\left[(j+2)(j+1+M_1)\right] =A_{j+1} \left[ (j + 1) +q_{_{2}} \right] + A_j \left[ (j + M_0-1)+ q_{_{1}}\right], \quad j \geq 0. \label{II.10}
\end{equation}
To solve this three-term recurrence relation, we use the conditions that for \(\forall j = n \geq 0\), we have \(A_{n+2} = 0\), \(A_{n+1} \neq 0\), and \(A_n \neq 0\). Consequently, the condition \(A_{n+1} \neq 0\) implies
\begin{equation}
q_{_{2}}=-(n+1)\Rightarrow\tilde{B}_{n,m} =\frac{ \left(n + |\tilde{m}| + \frac{3}{2}\right)}{2\tilde{m}}, \label{II.11}
\end{equation}
and \(A_{n} \neq 0\) yields
\begin{equation}
q_{_{1}}=-n(n+2|\tilde{m}|+2)\Rightarrow\tilde{\mathcal{E}}_{n,m} =  \left(n + |\tilde{m}| + \frac{1}{2}\right)\left(n + |\tilde{m}| + \frac{3}{2}\right). \label{II.12}
\end{equation}
These conditions indicate that the power series is truncated to a polynomial of order \(n + 1 \geq 0\). At this stage, it should be noted that the result in equation (\ref{II.11}) establishes a parametric relationship between the external magnetic field strength \(B_{\circ}\), the cosmological parameter $\Lambda$, and the cosmic string parameter \(0 < \alpha < 1\). On the other hand, equation (\ref{II.12}) provides the corresponding conditionally exact eigenvalues under these conditions.

We may now rewrite our three terms recurrence relation as
\begin{equation}
A_{j+2,n}=(j-n)\left(\frac{A_{j+1,n}+A_{j,n}(j+n+2|\tilde{m}|+2)}{(j+2)(j+2|\tilde{m}|+2)}\right), \quad j \geq 0. \label{II.10.1}
\end{equation}
Where, \(A_{0,n}=1\), and \(A_{1,n}=\frac{n+1}{2|\tilde{m}|+1}\). Therefore, our radial wave functions now read
\begin{gather}
R_{n,m}(\rho)= \mathcal{N}_{n,m} \, \sin(\rho)^{|\tilde{m}|} \sqrt{1 -\sin(\rho)} H_{n,m}(\rho),\nonumber  \\
H_{n,m}(\rho) = \sum_{j=0}^{n+1} A_{j,n} \,\left[-\sin(\rho)\right]^j, \label{II.13}
\end{gather}
where the coefficients of our polynomial of order \(n+1\) are given by (\ref{II.10.1}). Clearly, this radial wave function vanishes at \(\rho = 0, \pi\), satisfying the boundary conditions typically imposed in textbook treatments at the corresponding domain walls, which are generated by the magnetized BM-spacetime effective gravitational potential field. It could be interesting to know that equation (\ref{II.8}) is the general Heun equation and the power series \(H(\rho)\) in (\ref{II.9.1}) represents the general Heun function \(H_G(a, q, \alpha, \beta, \gamma, \delta, z)\) with:
\begin{gather}
z=-x,\,a = -1,\, q = -2\tilde{m}\tilde{B} + |\tilde{m}| + \frac{1}{2}, \, \alpha = |\tilde{m}| + 1 + \frac{\sqrt{4\tilde{\mathcal{E}} + 1}}{2}, \\
\beta = \frac{\left[( 3 + 4|\tilde{m}|)\sqrt{4\tilde{\mathcal{E}} + 1} + 2|\tilde{m}| - 8\tilde{\mathcal{E}} \right]}{2 + 4\sqrt{4\tilde{\mathcal{E}} + 1}}, \, \gamma = 2|\tilde{m}| + 1, \, \delta = \frac{1}{2}.
\end{gather}
Furthermore, equation (\ref{II.12}), together with equation (\ref{II.5}), implies:
\begin{equation}
\begin{split}
f(u)^2 E^2 = h(u)^2 \mathcal{K}_{nm} + m_{\circ}^2; \quad \tilde{B} = \frac{eB_{\circ}}{\sqrt{2\Lambda}}, \\
\mathcal{K}_{nm} = 2\Lambda \left( n + |\tilde{m}| + \frac{1}{2} \right)\left( n + |\tilde{m}| + \frac{3}{2} \right)+2\Lambda\tilde{B}^2 + k^2. \label{III.14}
\end{split}
\end{equation}
We will use this result in the subsequent analysis to explore the effects of RG corrections on the spectroscopic structure of KG particles and antiparticles in BM spacetime, subject to an external magnetic field.

\section{Rainbow gravity effects}\label{sec:3}

\begin{figure*}[ht!]
\centering
\includegraphics[width=0.3\textwidth]{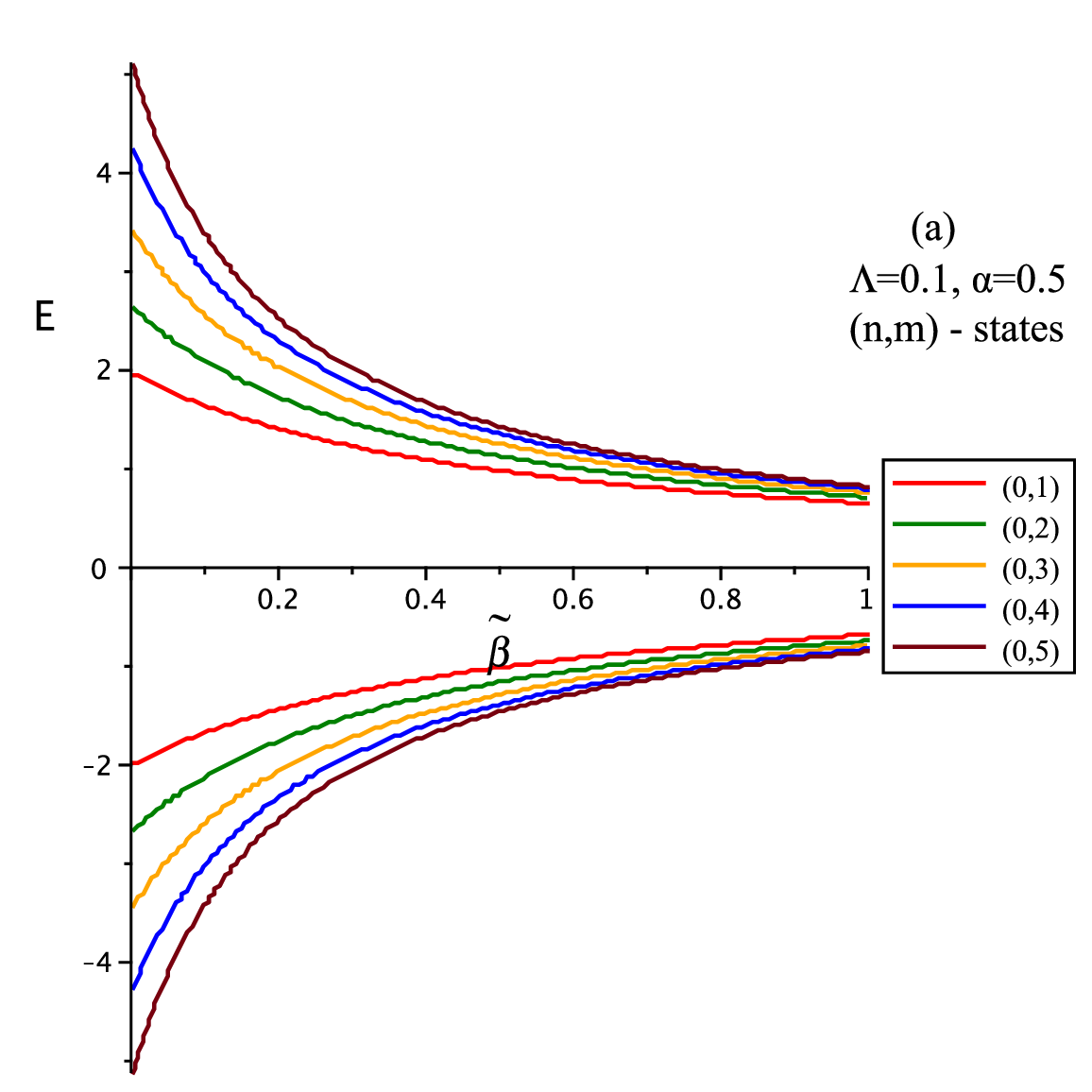}
\includegraphics[width=0.3\textwidth]{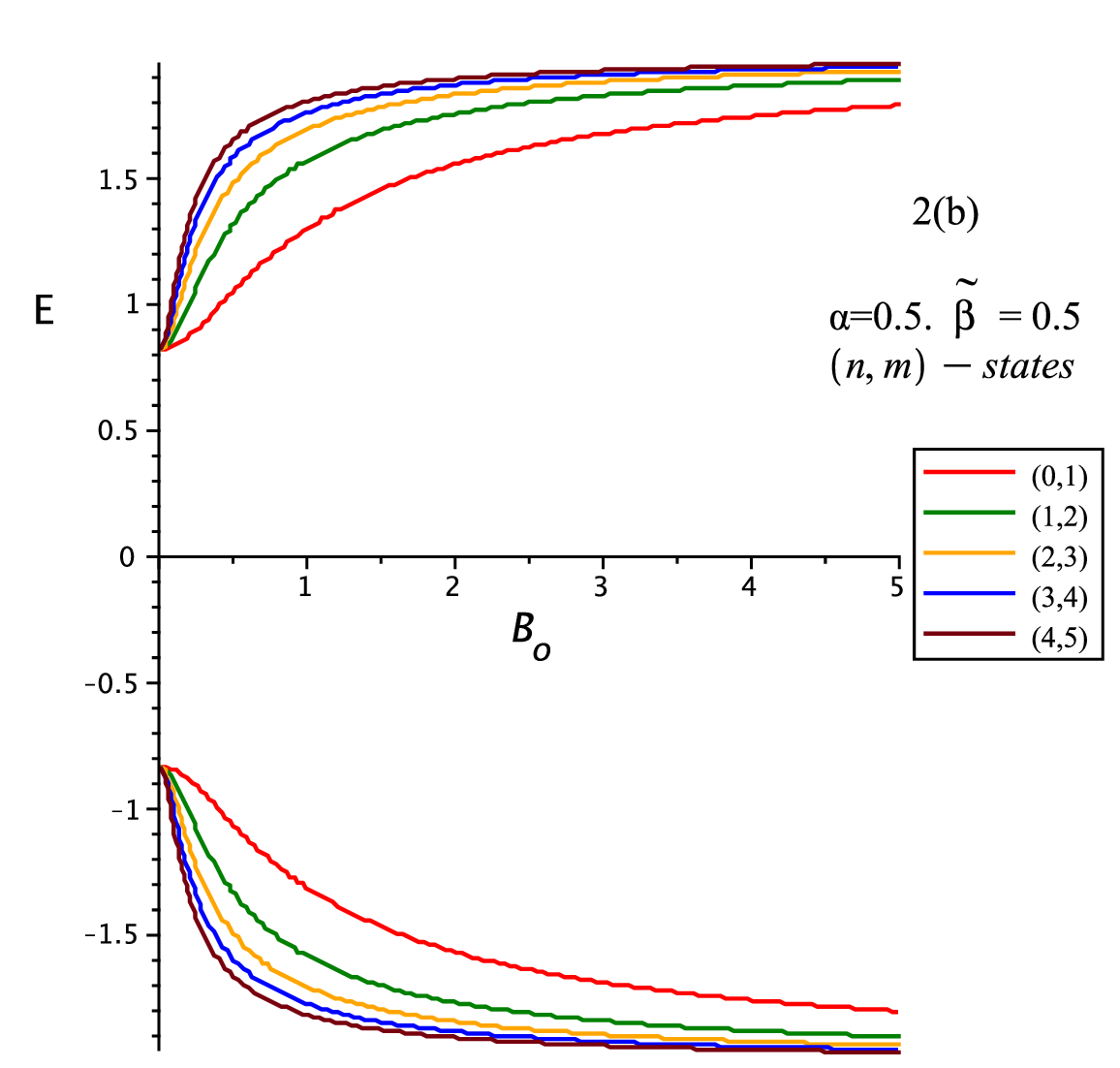}
\includegraphics[width=0.3\textwidth]{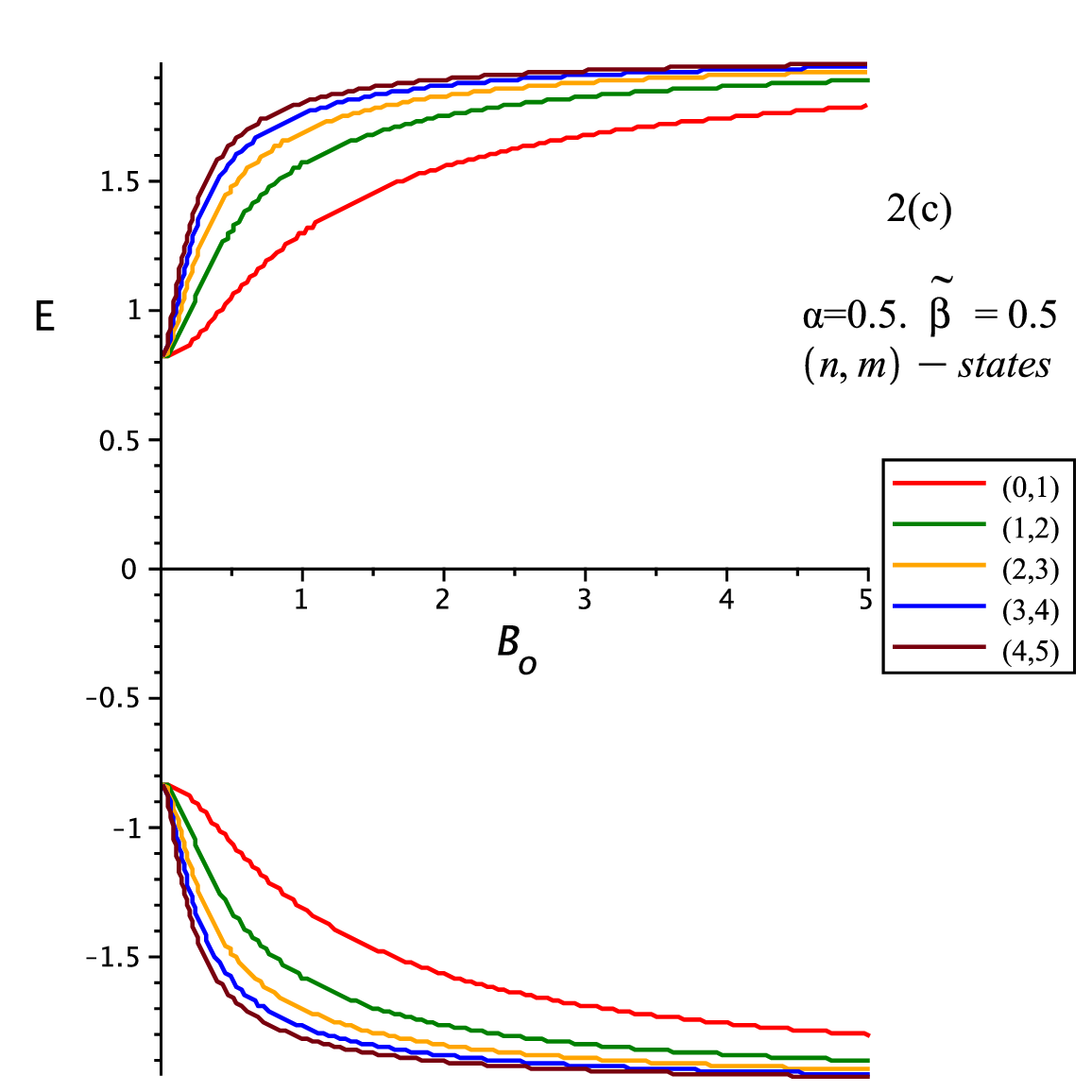}
\caption{\footnotesize  The figure shows the energy levels for KG particles and antiparticles given by (\ref{III.1.1}) so that we plot (a) $E$ against $\tilde{\beta}$ for $n=0$ and $m=1,2,3,4,5$, (b) $E$ against the magnetic field strength $B_\circ$ for $m=1,2,3,4,5$, and $n=0,1,2,3,4$, and (c) $E$ against $B_\circ$ for $n=0$ and $m=1,2,3,4,5$.}
\label{fig2}
\end{figure*}

In this section, we investigate the effects of RG on KG particles and antiparticles using two pairs of rainbow functions. The first pair, developed in the context of the varying speed of light hypothesis \cite{R1.36}, is given by: \(f\left( u \right) = (1 - \tilde{\beta}\left| E \right|)^{-1}, \, h\left( u \right) = 1\), where \( \tilde{\beta}=\beta /E_{p} \) is a parameter related to the energy scale. The second pair, motivated by loop quantum gravity (LQG) considerations \cite{R1.38,R1.39}, is expressed as: \( f\left( u \right) = 1, \, h\left( u \right) = \sqrt{1 - \tilde{\beta} \left| E \right|^\upsilon}, \, \upsilon = 1, 2\), with the energy scale \(u\) defined as \( 0 \leq u = |E| / E_p \leq 1 \). 

\subsection{Rainbow functions pair \(f(u) =\left( 1- \tilde{\beta}\left\vert E\right\vert \right) ^{-1}\), \(h(u) =1\)}

The substitution of the pair of rainbow functions \(f\left(u \right) = \left( 1 - \tilde{\beta}\left\vert E \right\vert \right)^{-1}\) and \(h\left( u \right) = 1\) into the result from (\ref{III.14}) leads to the following equation:
\begin{equation}
E^{2} = \mathcal{\tilde{K}}_{nm}\left( 1 - \tilde{\beta}\left\vert E \right\vert \right)^{2};\; \mathcal{\tilde{K}}_{nm} = \mathcal{K}_{nm} + m_{\circ}^{2}.
\label{III.1.1}
\end{equation}
It is important to observe that \(\left\vert E \right\vert = E_{+}\) for the test particles and \(\left\vert E \right\vert = -E_{-}\) for the antiparticles. Consequently, we obtain the following two equations: For the test particles, we have
\begin{equation}
E_{+}^{2}\left( 1 - \tilde{\beta}^{2} \mathcal{\tilde{K}}_{nm}\right) + 2 \tilde{\beta} \, \mathcal{\tilde{K}}_{nm} E_{+} - \mathcal{\tilde{K}}_{nm} = 0,
\label{III.1.2}
\end{equation}
and for the antiparticles, we obtain
\begin{equation}
E_{-}^{2} \left( 1 - \tilde{\beta}^{2} \mathcal{\tilde{K}}_{nm} \right) - 2 \tilde{\beta} \, \mathcal{\tilde{K}}_{nm} E_{-} - \mathcal{\tilde{K}}_{nm} = 0.
\label{III.1.3}
\end{equation}
Then the corresponding energies for both particles and antiparticles can be expressed as
\begin{equation}
E_{nm} = \pm \frac{\sqrt{\mathcal{\tilde{K}}_{nm}}}{1 + \tilde{\beta} \sqrt{\mathcal{\tilde{K}}_{nm}}}.
\label{III.1.4}
\end{equation}
It is clear that the energies for the KG particles and antiparticles are symmetric with respect to \(E = 0\) (documented in Fig. \ref{fig2}). However, it should be noted that the asymptotic behavior of \(\left\vert E_{\pm} \right\vert\) as $B_\circ \to \infty$ (consequently $\mathcal{\tilde{K}}_{nm}\to\infty\Rightarrow E_{nm}\to 1/\tilde{\beta}$ ) converges to \(\left\vert E_{\pm} \right\vert \sim 2 = 1/\tilde{\beta} = E_{p}/\beta\), as clearly observed in Figs. \ref{fig2}(b) and \ref{fig2}(c). As a result, it follows that \(\left\vert E_{\pm} \right\vert \leq E_{p} \Longrightarrow \left\vert E_{\pm} \right\vert_{\max} = E_{p}\) for the case when \(\beta_{\min} = 1\). 
\begin{figure*}[ht!]
\centering
\includegraphics[width=0.3\textwidth]{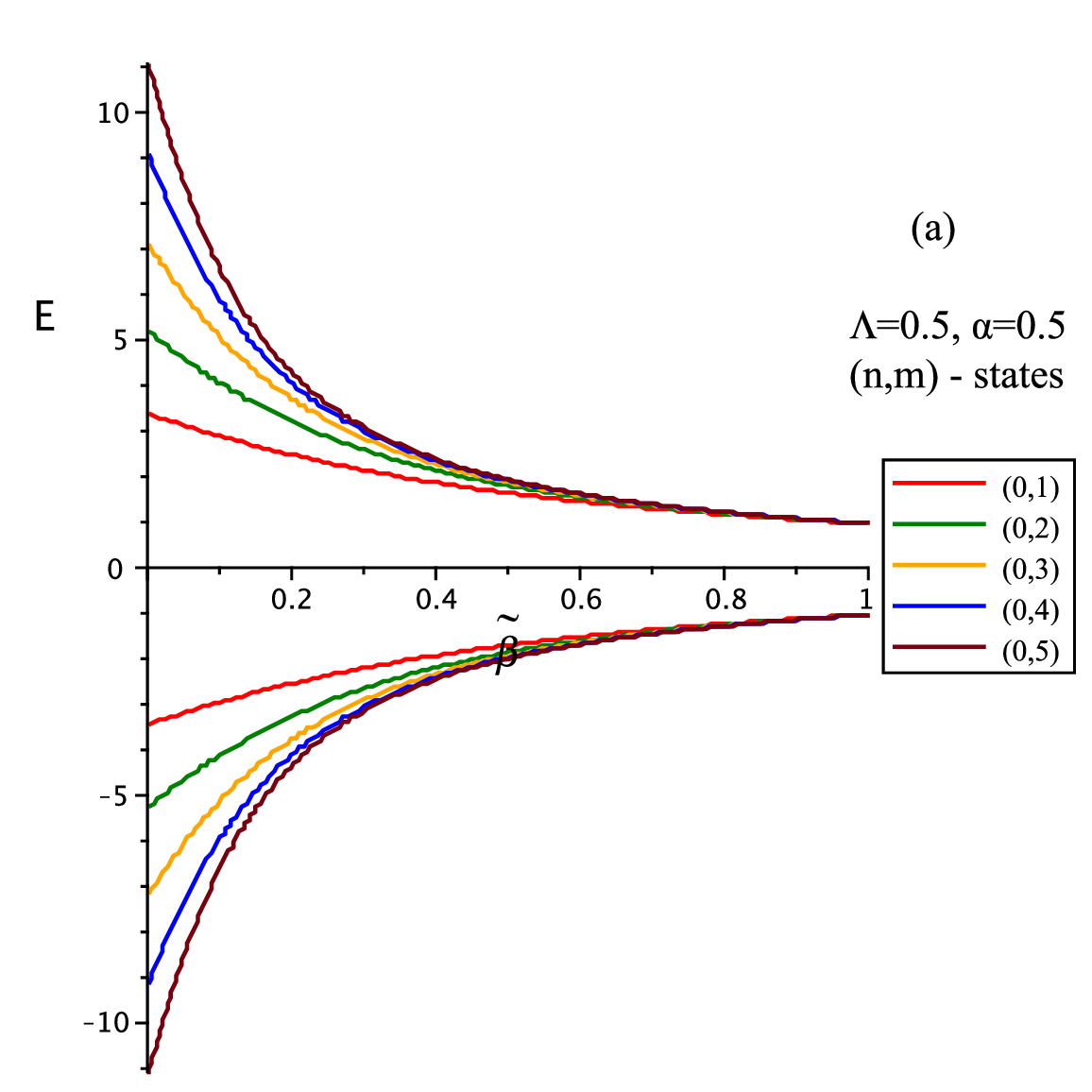}
\includegraphics[width=0.3\textwidth]{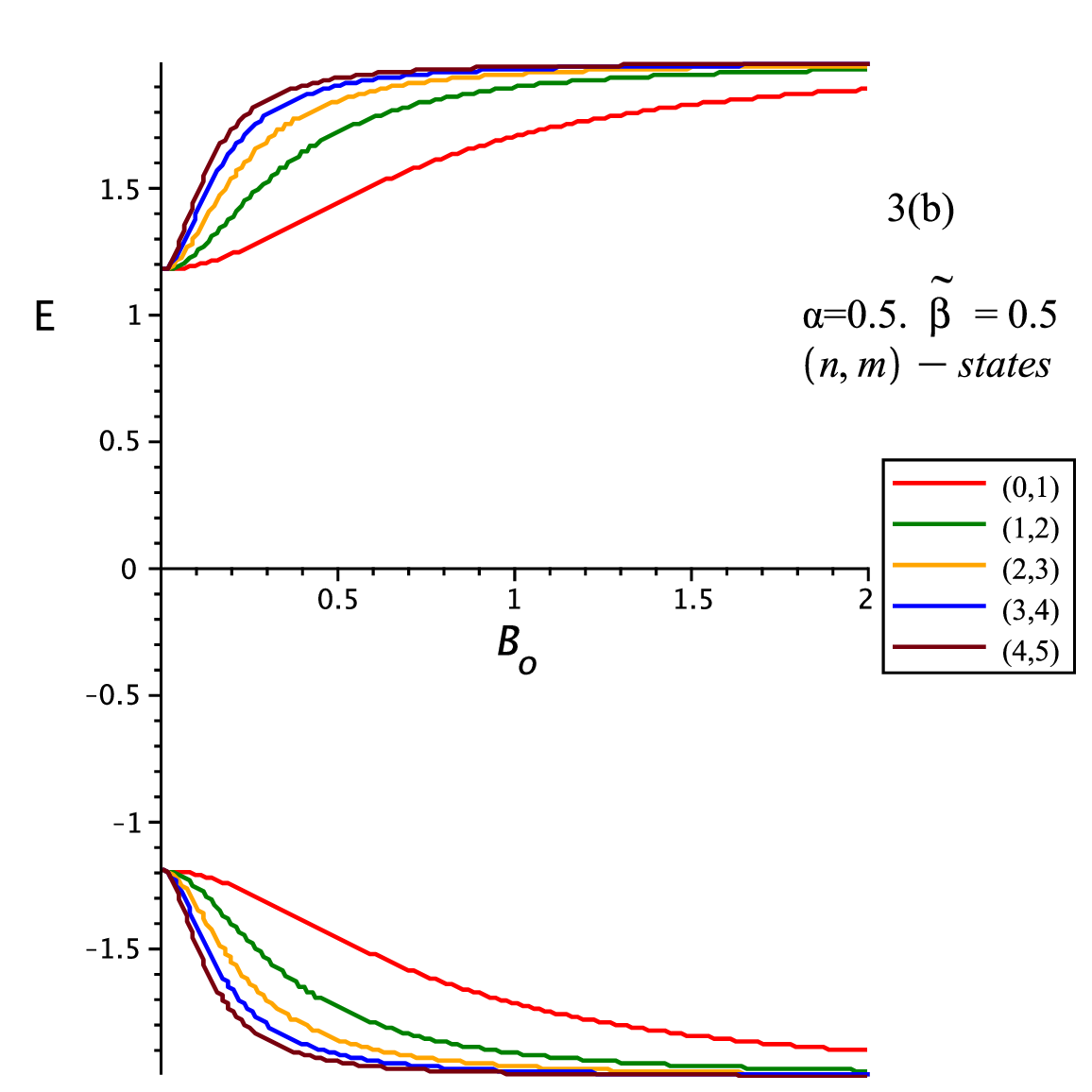}
\includegraphics[width=0.3\textwidth]{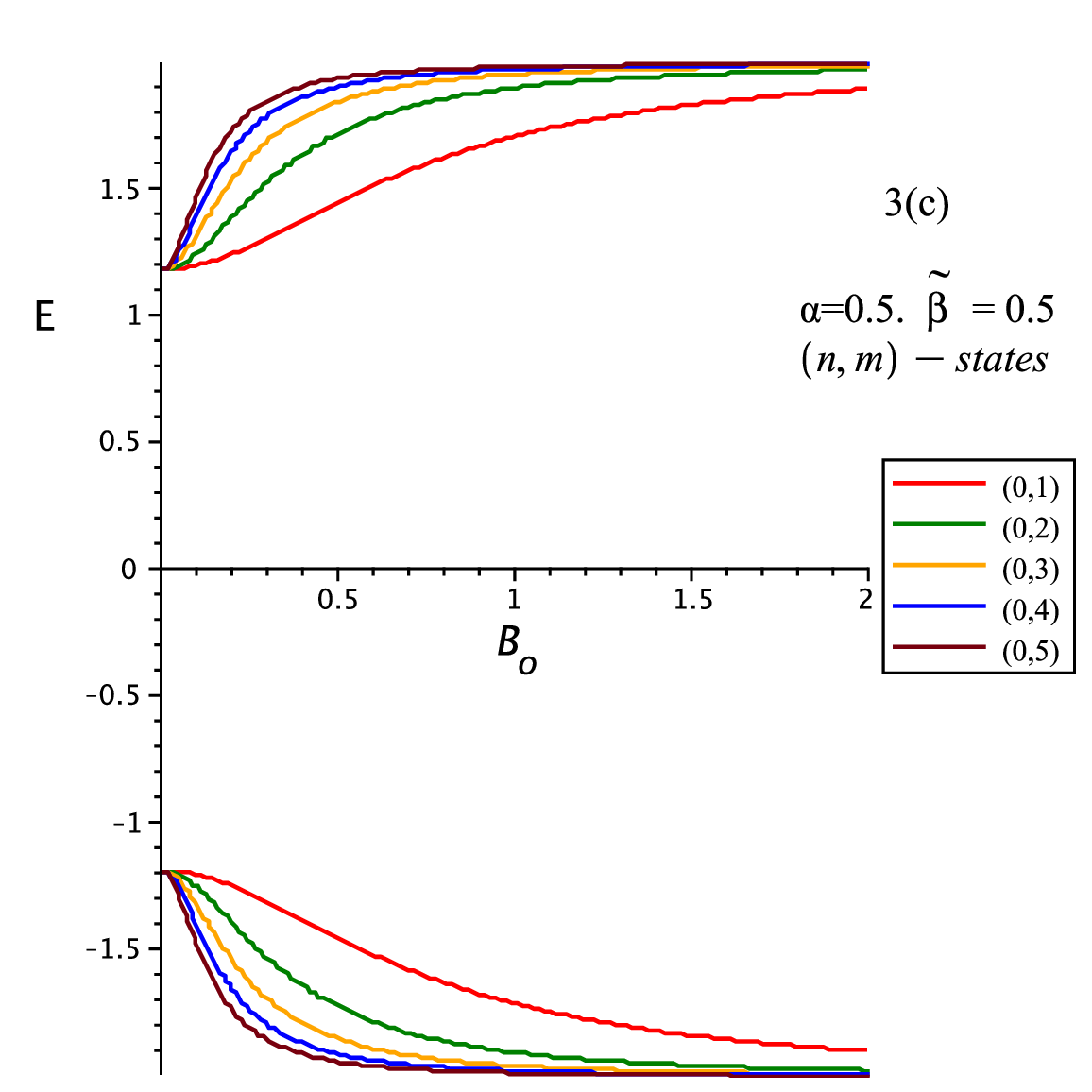}
\caption{\footnotesize  The figure shows the energy levels for KG particles and antiparticles given by (\ref{III.3.1}), and we plot (a) $E$ against $\tilde{\beta}$ for $n=0$ and $m=1,2,3,4,5$, (b) $E$ against the magnetic field strength $B_\circ$ for $n=0,1,2,3,4$ and $m=1,2,3,4,5$, and (c) $E$ against $B_\circ$ for $n=0$ and $m=1,2,3,4,5$.}
\label{fig3}
\end{figure*}
\begin{figure*}[ht!]
\centering
\includegraphics[width=0.3\textwidth]{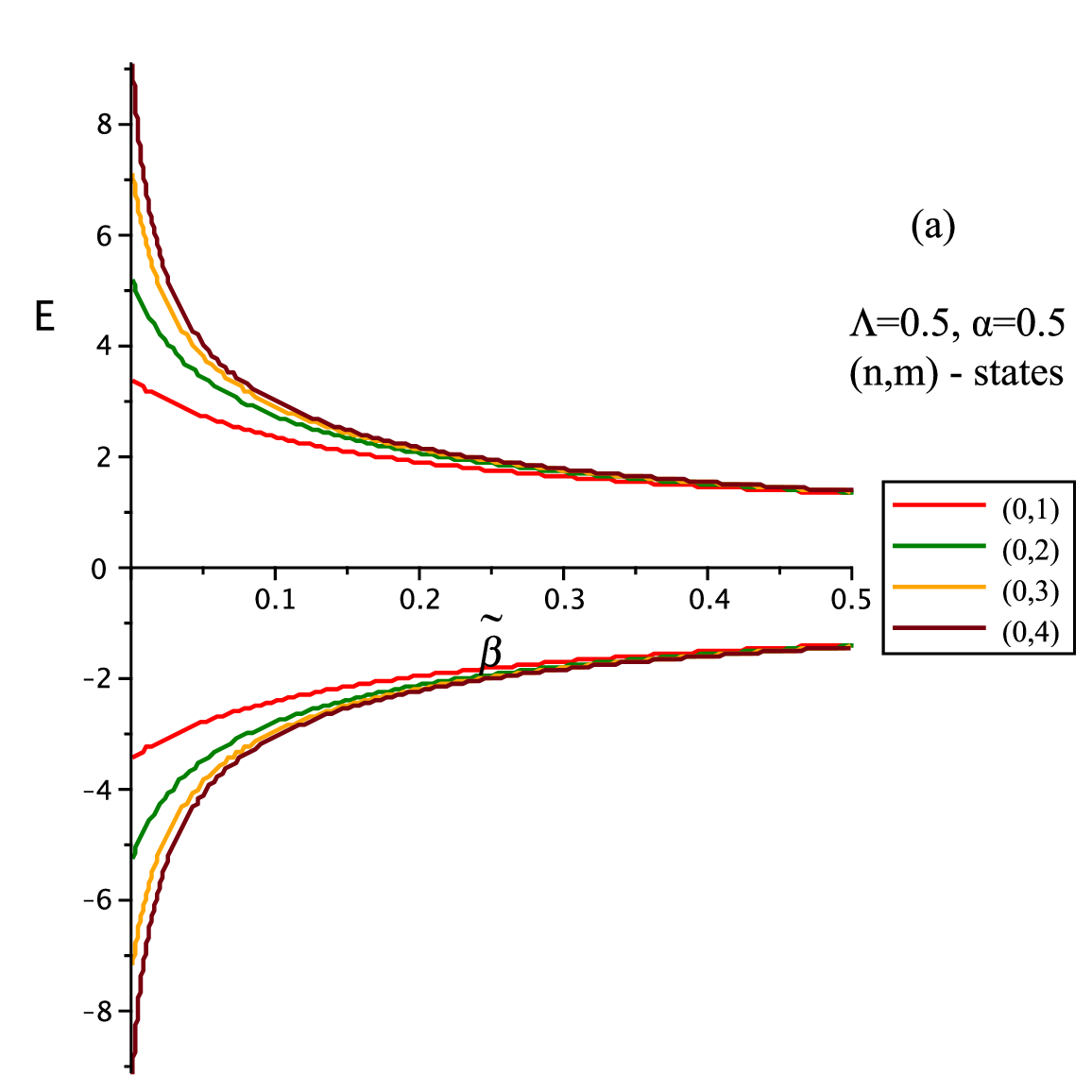}
\includegraphics[width=0.3\textwidth]{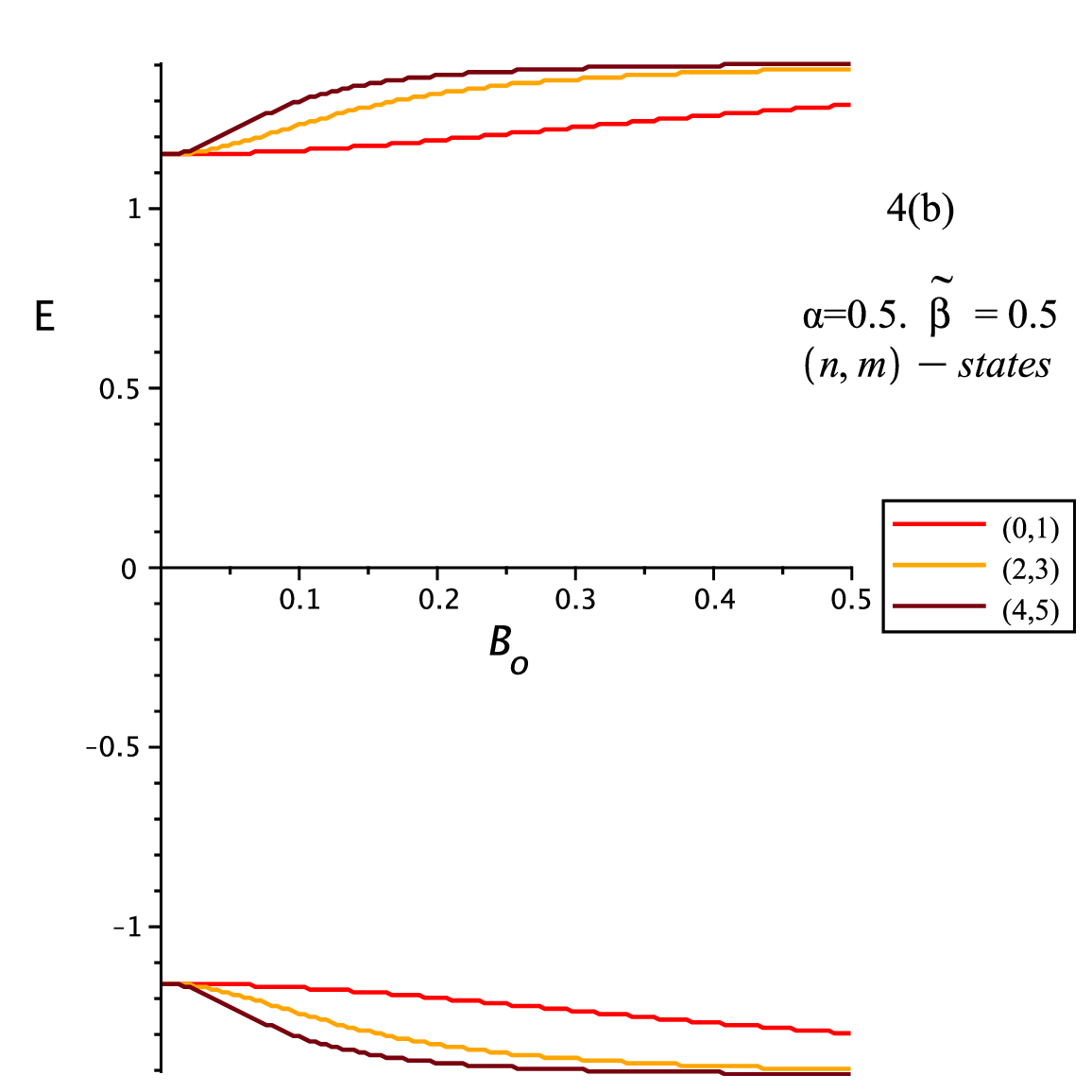}
\includegraphics[width=0.3\textwidth]{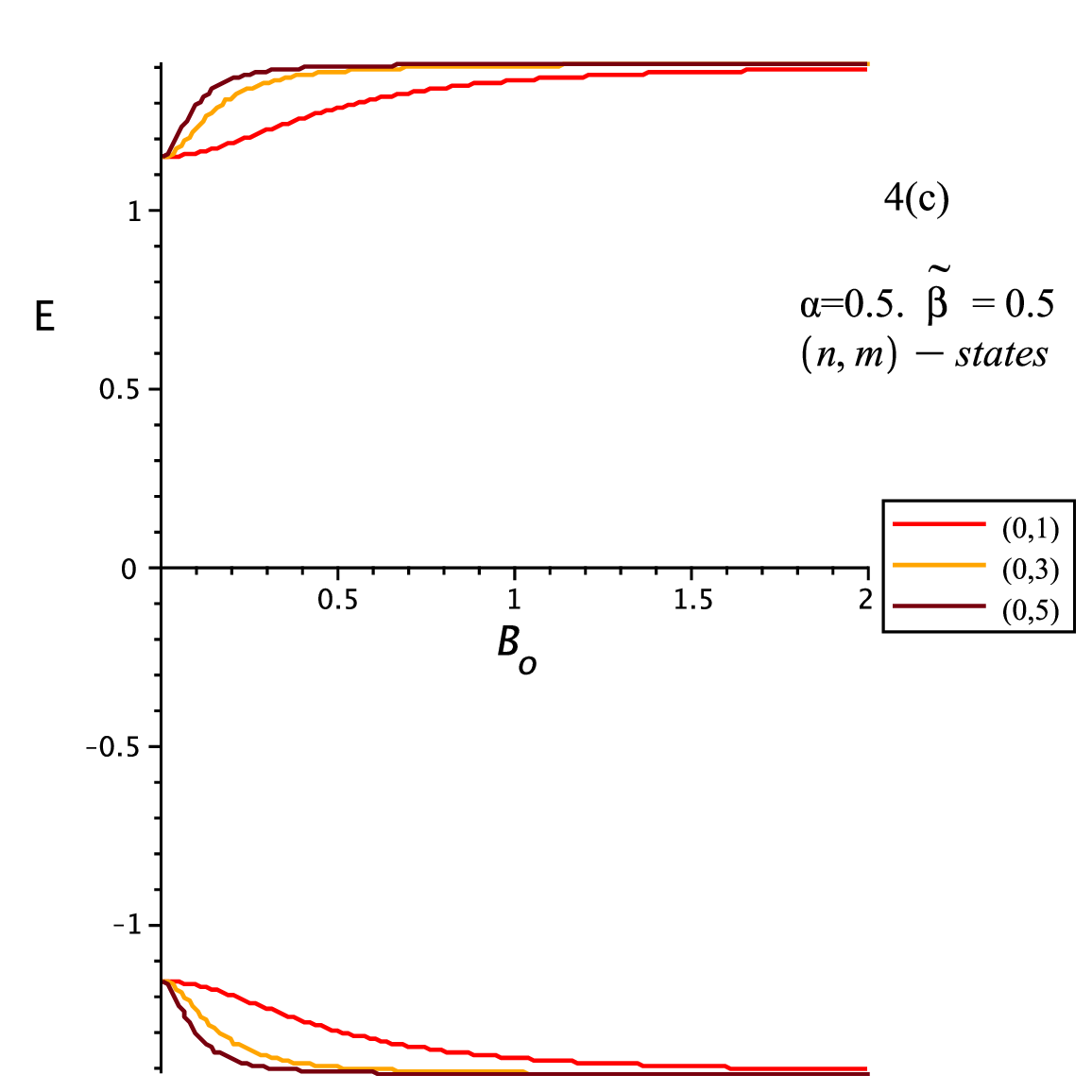}
\caption{\footnotesize The figure shows the energy levels for KG particles and antiparticles given by (\ref{III.3.1}), so that we plot (a) \(E\) against \(\tilde{\beta}\) for \(n=0\) and \(m=1,2,3,4\), (b) \(E\) against the magnetic field strength \(B_\circ\) for \(n=0,2,4\) and \(m=1,3,5\), and (c) \(E\) against \(B_\circ\) for \(n=0\) and \(m=1,3,5\).}
\label{fig4}
\end{figure*}

\subsection{Rainbow functions pair $f(u)=1$, $h\left(u \right) = \sqrt{1-\tilde{\beta}\left\vert E\right\vert^\upsilon}$; $\upsilon=1,2$ }

Such pairs of rainbow functions are motivated by loop quantum gravity \cite{R1.38,R1.39}, which have been found to fully comply with the RG principles, ensuring that the Planck energy \( E_p \) (which represents the maximum achievable energy for both particles and antiparticles, for example, \cite{R1.34,R1.35,R1.40}). Therefore, it is of particular interest to investigate the behavior of these pairs when applied to KG particles and antiparticles in a magnetized BM spacetime. We begin with the parameter \(\upsilon = 1\), which yields, using (\ref{III.14}), the following expression:
\begin{equation}
    E^2 + \tilde{\beta} \mathcal{K}_{nm} |E| - \tilde{\mathcal{K}}_{nm} = 0. \label{III.3.1}
\end{equation}
From this, the energy solutions are derived as follows:
\begin{equation}
    E_+ = \frac{-\tilde{\beta} \mathcal{K}_{nm}}{2} + \frac{1}{2} \sqrt{\tilde{\beta}^2 \mathcal{K}_{nm}^2 + 4 \tilde{\mathcal{K}}_{nm}}, \label{III.3.2}
\end{equation}
and
\begin{equation}
    E_- = \frac{\tilde{\beta} \mathcal{K}_{nm}}{2} - \frac{1}{2} \sqrt{\tilde{\beta}^2 \mathcal{K}_{nm}^2 + 4 \tilde{\mathcal{K}}_{nm}}. \label{III.3.3}
\end{equation}
These energy levels are visualized in Figure \ref{fig3}. It is evident from (\ref{III.3.1}) that as $B_\circ \to \infty$ (hence $\mathcal{\tilde{K}}_{nm}\to\infty\Rightarrow |E|\to 1/\tilde{\beta}$ ) the maximum energy, \( |E|_{\text{max}} \), approaches \( 1/\tilde{\beta} = 2 \). This behavior illustrates the asymptotic convergence of the energies of KG particles and antiparticles in the BM-RG spacetime tend to approach \( |E|_{\text{max}}=E_p \) , as clearly observed in Figure \ref{fig3}. 

\vspace{0.15cm}
\setlength{\parindent}{0pt}

Next, we consider the case where \(\upsilon = 2\), which leads to the following equation:
\begin{equation}
    E^2 = \frac{\tilde{\mathcal{K}}_{nm}}{1 + \tilde{\beta} \mathcal{K}_{nm}} \quad \Rightarrow \quad E_\pm = \pm \sqrt{\frac{\tilde{\mathcal{K}}_{nm}}{1 + \tilde{\beta} \mathcal{K}_{nm}}}. \label{III.3.4}
\end{equation}
The corresponding energy levels are plotted in Figure \ref{fig4}. This plot further confirms the asymptotic behavior of the energy levels as $B_\circ \to \infty$ (subsequently $\mathcal{\tilde{K}}_{nm}\to\infty\Rightarrow E_{nm}\to \sqrt{1/\tilde{\beta}}$ ), which, as indicated by (\ref{III.3.4}), implies that \( |E|_{\text{max}} \sim \sqrt{1/\tilde{\beta}} = \sqrt{2} \approx 1.41 \). This trend is consistent with the analytical result derived from (\ref{III.3.4}) and is also shown in Figure \ref{fig4}.

\section{Concluding remarks }\label{sec:4}

In this paper, we have studied the effects of rainbow gravity on the spectroscopic structure of KG-particles/antiparticles in an external magnetic field in the Bonnor-Melvin spacetime background. In the process, we emphasized that such a magnetized spacetime fabric generates infinite domain walls at some well-pronounced values of the radial coordinate, namely, at \(\rho=\sqrt{2\Lambda}r=0,\pi,2\pi,\cdots\).  This, in turn, would suggest that the radial range within which a quantum particle moves is \(\rho \in [0,\pi]\). In fact, this would not only agree with \v{Z}ofka's \cite{R1.1} suggestion that \(\rho = \sqrt{2\Lambda}r = \pi\) is the location of an axis of some sort, but would also clearly identify this axis as a domain wall, one of the well-known topological defects in the early universe. Consequently, this would, in fact, provide brute-force evidence that the asymptotic approximation \(r \ll 1 \Rightarrow \sin(r) \sim r \Rightarrow \tan(r) \sim r\) (e.g., \cite{R1.4,R1.4.1}) is rendered inappropriate in this context. Adopting such an asymptotic tendency would not only change the dynamics of the quantum mechanical system but also dismiss the domain walls, which are the very fundamental characteristic of the magnetized BM-spacetime, 
as clearly illustrated in Figures \ref{fig1}(a) and \ref{fig1}(b).

\vspace{0.15cm}
\setlength{\parindent}{0pt}

Under such domain walls settings, we have studied the effects of RG on KG particles/antiparticles in BM-spacetime and in an external magnetic field. In so doing, we have used two pairs of rainbow functions. The first pair, which is developed in the context of the hypothesis of varying light speed \cite{R1.36}, is given by: \(f\left( u \right) =  (1 - \tilde{\beta} |E|)^{-1}, \, h\left( u \right) = 1\). The second pair, on the other hand, is motivated by loop quantum gravity (LQG) \cite{R1.38,R1.39}, and reads \( f\left( u \right) = 1, \, h\left( u \right) = \sqrt{1 - \tilde{\beta} \left| E \right|^\upsilon}, \, \upsilon = 1, 2\), where \( 0 \leq u =| E| / E_p \leq 1 \). We have observed that the two sets of rainbow functions have complied well with RG theory, so that they have secured the Planck energy \( E_p \) as the maximum possible energy for both particles and antiparticles, in agreement with previous works in the literature (e.g., \cite{R1.34, R1.35, R1.40}).
\begin{figure*}[ht!]
\centering
\includegraphics[width=0.3\textwidth]{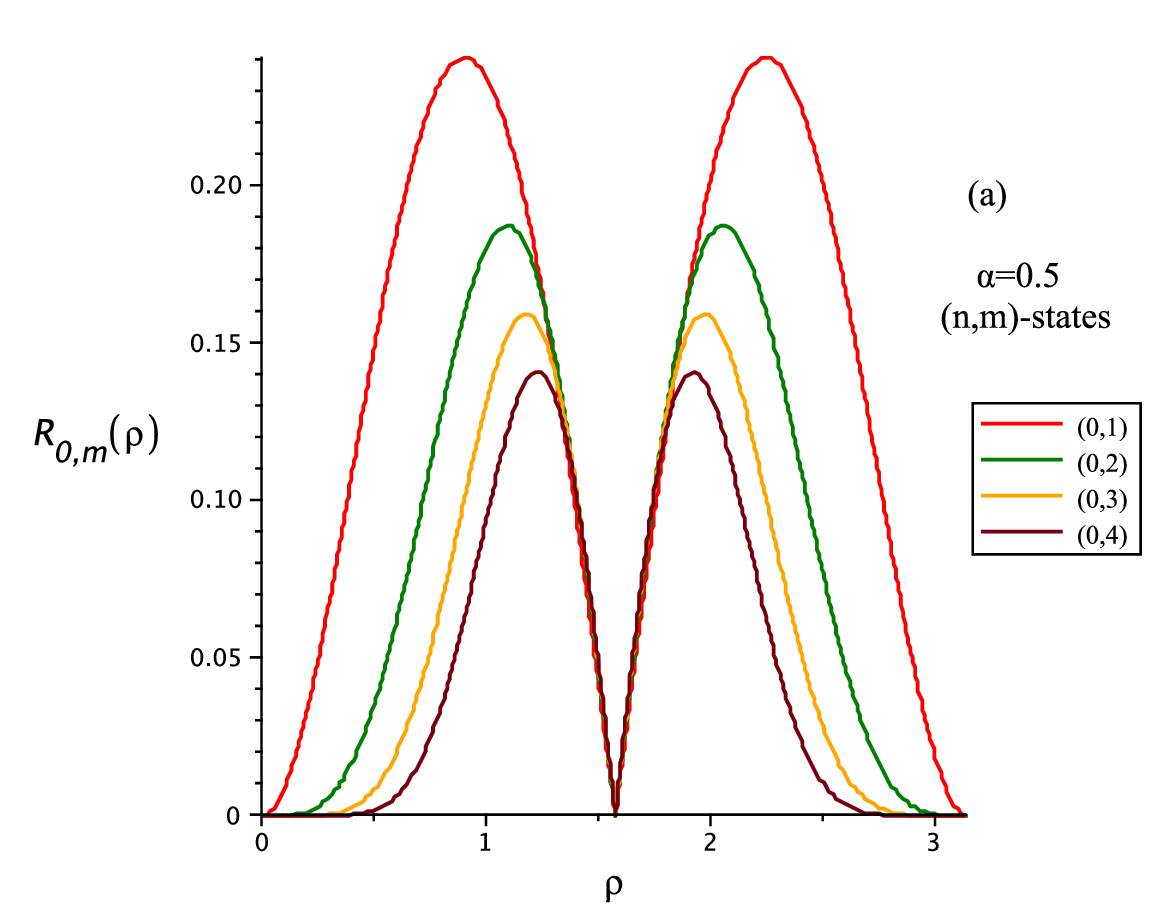}
\includegraphics[width=0.3\textwidth]{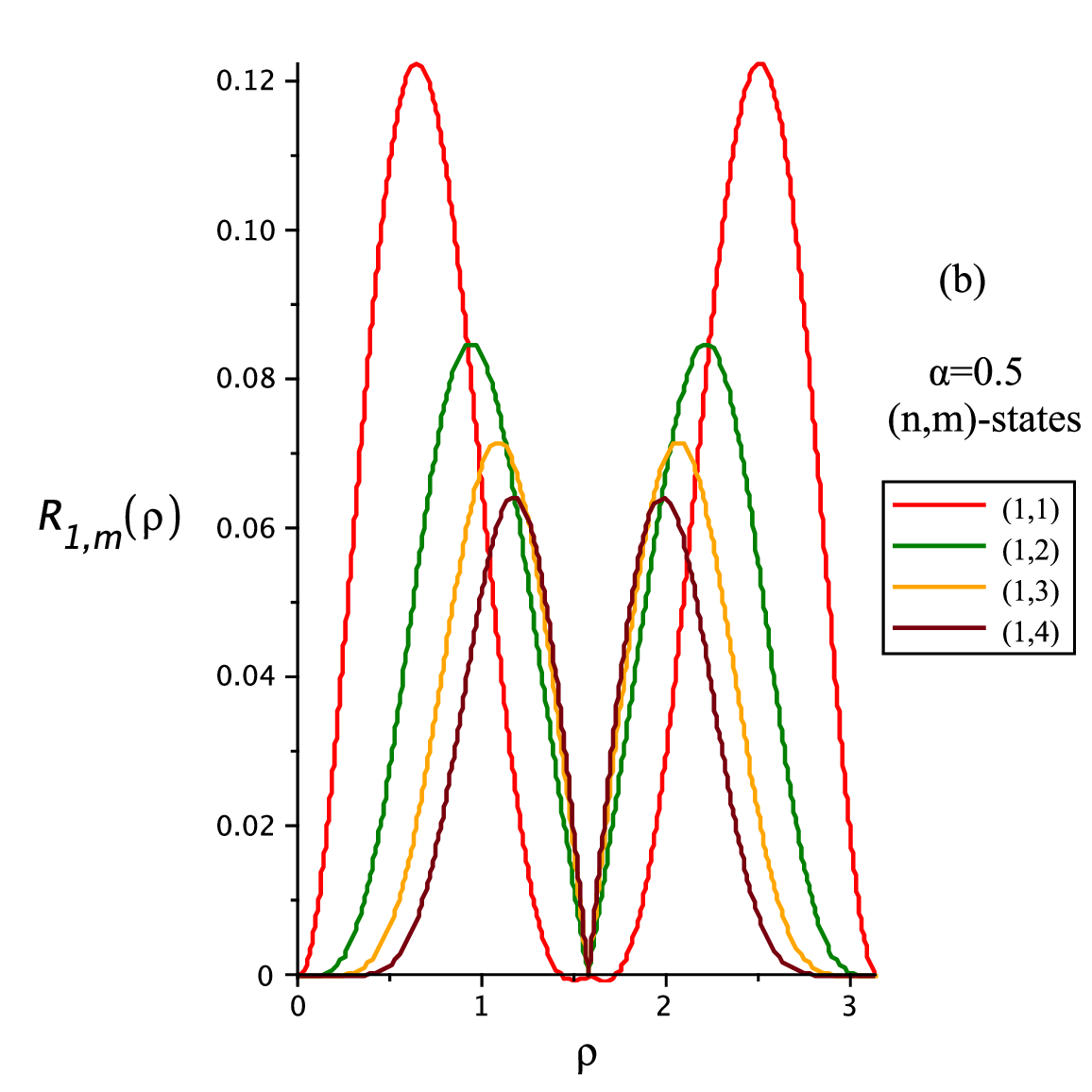}
\includegraphics[width=0.3\textwidth]{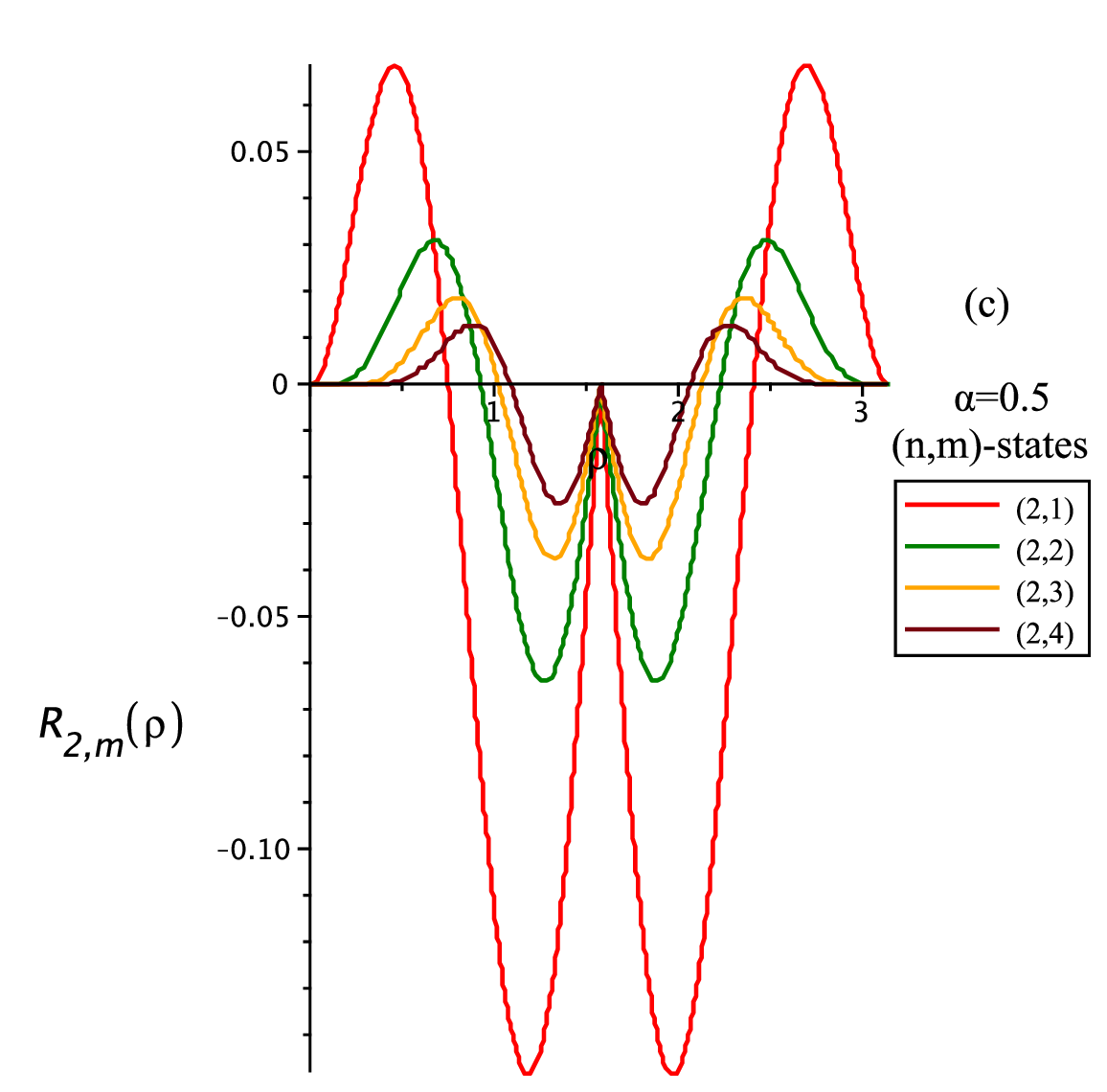}
\caption{\footnotesize The figure shows the first three radial wave functions of (\ref{II.13}) so that we have in (a) \(R_{0,m}(\rho)\), in (b) \(R_{1,m}(\rho)\), and in (c) \(R_{2,m}(\rho)\).}
\label{fig5}
\end{figure*}
\begin{figure*}[ht!]
\centering
\includegraphics[width=1\textwidth]{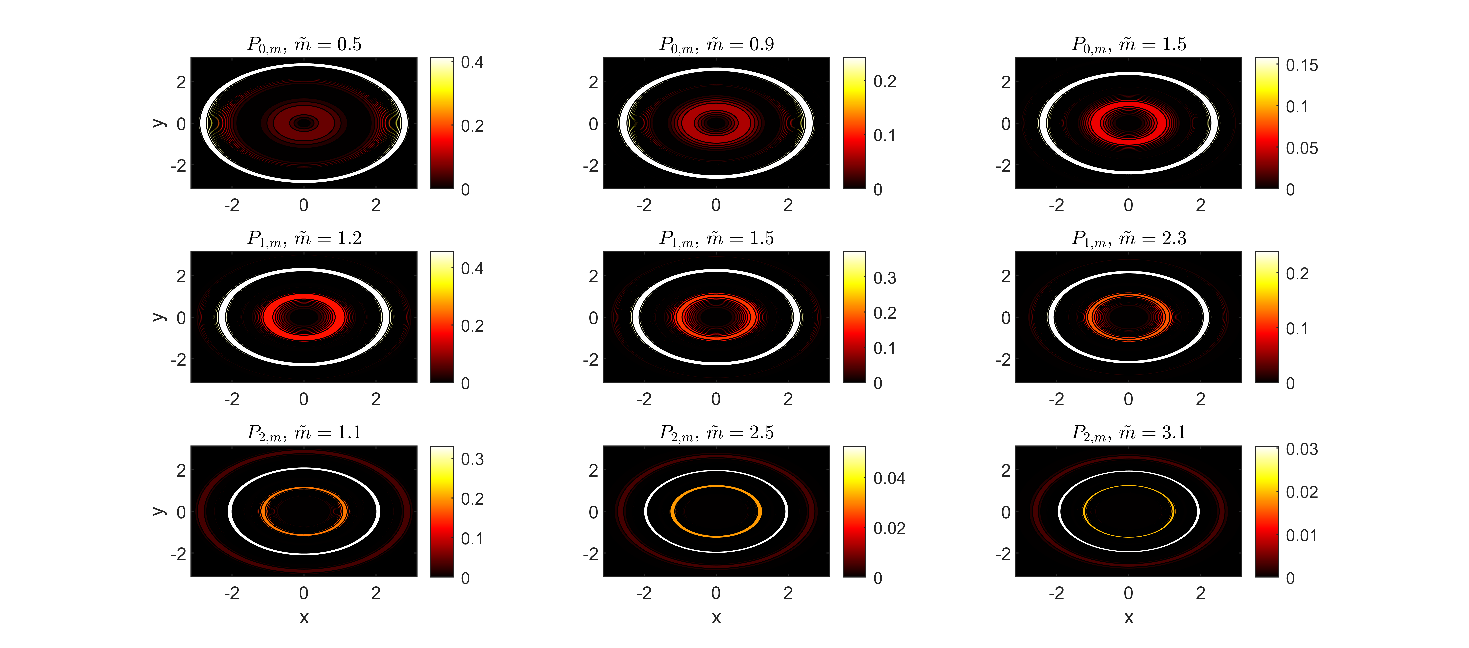}
\caption{\footnotesize Radial Probability Densities \(P_{0,m}\), \(P_{1,m}\), and \(P_{2,m}\) for varying values of \(\tilde{m}\). The figure presents the radial probability densities \(P_{0,m}\), \(P_{1,m}\), and \(P_{2,m}\) in the \(x\)-\(y\) plane for \(\tilde{m} = 0.5, 0.9, 1.5\) (top row), \(\tilde{m} = 1.2, 1.5, 2.3\) (middle row), and \(\tilde{m} = 1.1, 2.5, 3.1\) (bottom row). Each plot depicts the spatial probability distribution for different quantum states, emphasizing the effect of \(\tilde{m}\) on the radial density within the range \(\rho \in [0, \pi]\).}
\label{fig:new:1}
\end{figure*}

\vspace{0.15cm}
\setlength{\parindent}{0pt}

Moreover, in connection with the radial wave function reported in (\ref{II.13}) as
\begin{gather}
R_{n,m}(\rho)= \mathcal{N}_{n,m} \, \sin(\rho)^{|\tilde{m}|} \sqrt{1 -\sin(\rho)} H_{n,m}(\rho),  \\
H_{n,m}(\rho) = \sum_{j=0}^{n+1} A_{j,n} \,[-\sin(\rho)]^j, \label{IV.1.1}
\end{gather}
where \(A_{0,n}=1,\, A_{1,n}=\frac{n+1}{2|\tilde{m}|+1}\), and \(A_{j,n}\, 's\) are given by (\ref{II.10.1}) so that
\begin{equation}
\begin{split}
   A_{2,n}&=-n\frac{ \left[A_{1,n}+ A_{0,n}(n+2|\tilde{m}|+2)\right]}{2(2|\tilde{m}|+2)}, \\
   A_{3,n}&=-(n-1)\frac{ \left[A_{2,n}+ A_{1,n}(n+2|\tilde{m}|+3)\right]}{3(2|\tilde{m}|+3)},\\
   A_{4,n}&=-(n-2)\frac{ \left[A_{3,n}+ A_{2,n}(n+2|\tilde{m}|+4)\right]}{4(2|\tilde{m}|+4)}.
     \end{split}
\end{equation}
In Figure \ref{fig5}, we plot the first three radial wave functions. They all vanish at the domain walls at \(\rho=0,\pi\) as a textbook tendency of the behavior of the wave functions at the infinite domain walls. More interestingly, it is obvious that the very existence of domain walls (which mandates the validity of \(\sin(\rho)\in[0,1]\) and not \(\sin(\rho)\in[-1,1]\)) prevents the sinusoidal radial wave function from performing a complete cycle, i.e,.  \(\rho\notin[0,2\pi]\) but rather \(\rho\in[0,\pi]\). Moreover, the term \( \sqrt{1 -\sin(\rho)}\) in the radial wave function leaves a brutal signature at \(\rho=\pi/2\) so that at this value the radial wave function collapses to zero. 

\vspace{0.15cm}
\setlength{\parindent}{0pt}

On the other hand, we plot, in Figure \ref{fig:new:1}, the radial probability density function \( P_{n,m}(\rho) = \int_{0}^{\rho} |R_{n,m}(\rho)|^2 \rho \,  d\rho \) for the first three states \(n=0,1,2\) at different values for \(\tilde{m}\). In doing so, we used the transformations \( x = \rho \cos{\varphi} \) and \( y = \rho \sin{\varphi} \) to visualize the radial probability densities as functions of position in two dimensions. This figure reveals that the modes are restricted to rotating, ring-shaped structures (note that \( |m| \neq 0 \)), or, in other words, the bosonic states manifest themselves as spinning vortices. It is clear that the out-of-plane magnetic field plays a pivotal role in modulating both the structure and intensity of the spinning magnetized vortex configurations associated with scalar bosonic states. Physically, this magnetic field introduces a form of rotational confinement that alters the spatial distribution of the probability densities, thereby dictating the localization behavior of KG bosons within the region enclosed by the magnetized BM domain walls. As illustrated in Figure~\ref{fig:new:1}, the shape and amplitude of these probability densities exhibit a strong dependence on quantum numbers, particularly the quantum number \(n\) and the topologically modified magnetic quantum number \(\tilde{m}\). The latter encodes both the magnetic properties and the topological effects induced by the geometry of the magnetized BM spacetime. For ground state configurations (\(n = 0\)) with varying \(\tilde{m}\), the probability density profiles tend to be broad and smooth, signifying a relatively weaker localization and a more dispersed spatial distribution of the KG bosons. In contrast, excited states (\(n > 0\)) are characterized by sharper and more localized peaks in their probability densities, indicating enhanced confinement due to stronger magnetic interactions and geometric constraints. Consequently, the excited bosonic states form increasingly compact and tightly wound spinning, magnetized vortices. These findings underscore the profound influence of both the external magnetic field strength and the underlying spacetime geometry on the physical characteristics of the bosonic modes. The quantized external magnetic field not only alters the energy spectrum, but also imprints distinct spatial signatures on the wavefunctions, signatures that are governed by the excitation level and the topological features encapsulated in \(\tilde{m}\). Moreover, our results strongly indicate that the magnetic background considered here can facilitate the emergence of rotating and dynamically evolving matter rings, regardless of whether the KG particles and antiparticles are massive \(m_{\circ} \neq 0\) or massless \(m_{\circ} = 0\).

\section*{Credit authorship contribution statement}

\textbf{Omar Mustafa}: Conceptualization, Methodology, Formal Analysis, Writing – Original Draft, Investigation, Visualization, Writing – Review and Editing, Project Administration.\\
\textbf{Abdullah Guvendi}: Conceptualization, Methodology, Formal Analysis, Writing - Original Draft, Investigation, Writing - Review, and Editing. 

\section*{Data availability}

This manuscript has no associated data.

\section*{Conflicts of interest statement}

No conflict of interest declared by the authors.

\section*{Funding}

No funding has been received for this study.

\end{document}